\documentclass[12pt,epsf,epsfig,psfig]{article}
\usepackage{graphicx}
\usepackage{epstopdf}
\usepackage{epsfig}
\usepackage{cite}
\def\be{\begin{equation}}
\def\ee{\end{equation}}

\def\lsim{\raise0.3ex\hbox{$<$\kern-0.75em\raise-1.1ex\hbox{$\sim$}}}
\def\gsim{\raise0.3ex\hbox{$>$\kern-0.75em\raise-1.1ex\hbox{$\sim$}}}

\def\NP{{ Nucl.\ Phys.\ }}
\def\PL{{ Phys.\ Lett.\ }}
\def\PR{{ Phys.\ Rev.\ }}

\def\PRL{{ Phys.\ Rev.\ Lett.\ }}

\def\EP{{ Europ.\ Phys.\ J.\ C}}

\begin{document}



\hfill Bielefeld, 6.\ 9.\ 2017

\vskip1cm

\centerline{\Large \bf Strangeness Production and Color Deconfinement}

\vskip0.6cm 

\centerline{\bf P.\ Castorina$^{\rm a,b}$, S.\ Plumari$^{\rm a,c}$
and H.\ Satz$^{\rm d}$} 

\bigskip

\centerline{a: Dipartimento di Fisica ed Astronomia, 
Universit\'a di Catania, Italy}

\centerline{b: INFN, Sezione di Catania, Catania, Italy}

\centerline{c: INFN-LNS, Catania, Italy} 

\centerline{d: Fakult\"at f\"ur Physik, Universit\"at Bielefeld, Germany}

\vskip1cm

\centerline{\large \bf Abstract:}

\bigskip

The relative multiplicities for hadron production in different high energy 
collisions are in general well described by an ideal gas of all hadronic 
resonances, except that under certain conditions, strange particle rates 
are systematically reduced. We show that the suppression factor $\gamma_s$, 
accounting for reduced strange particle rates in $pp$, $pA$ and $AA$ 
collisions at different collision energies, becomes a universal function when 
expressed in terms of the initial entropy density $s_0$ or the initial
temperature $T$ of the produced thermal
medium. It is found that $\gamma_s$ increases from about 0.5 to 
1.0 in a narrow temperature range around the quark-hadron transition 
temperature $T_c \simeq 160$ MeV. Strangeness suppression thus disappears
with the onset of color deconfinement; subsequently, full equilibrium 
resonance gas behavior is attained.

\vskip1cm

The relation of color deconfinement and strange hadron production in high
energy collisions has attracted much attention over more than thirty years.
It had been observed that in elementary ($pp,~e^+e^-$) collisions as well
as in low energy nucleus-nucleus interactions, the production of strange
particles was suppressed relative to the rates expected for an equilibrium
resonance gas. In this context, it was argued early on that quark gluon 
plasma formation would enhance strange particle production in nucleus-nucleus 
collisions \cite{raf,raf-mul}, bringing the produced system closer to 
equilibrium resonance gas rates. Over the years, different mechanisms for 
such an effect were discussed \cite{HRT,PBM,CS13,CS16,CPS}. The aim of the 
present paper is to provide clear experimental evidence showing that in 
$AA,~pA$ and $pp$ collisions, strangeness production as function of the 
initial temperature of the produced thermal medium follows a universal form, 
with suppression ending in a narrow temperature band around the color 
deconfinement temperature. 
 
\medskip

High energy proton-proton, proton-nucleus and nucleus-nucleus collisions lead
to abundant multihadron production. The relative rates of the secondaries thus
produced are well accounted for by an ideal gas of all hadrons and hadronic
resonances at fixed temperature $T$ and baryochemical potential $\mu$, with 
one well-known caveat. Strangeness production is reduced with respect to the 
rates thus predicted. This suppression can, however, be taken into account by
one further parameter, $0 < \gamma_s \leq 1$, if the predicted rate for 
a hadron species containing $\nu=1,2,3$ strange quarks is suppressed by 
the factor $\gamma_s^{\nu}$ \cite{LRT}. 

\medskip 

The basic quantity for the resonance gas description is the grand-canonical
partition function for an ideal gas at temperature $T$ in a spatial volume $V$
\be
\ln Z(T) = V \sum_i {d_i \gamma_s^{\nu_i}\over (2\pi)^3}~\! \phi(m_i,T),
\ee
with $d_i$ specifying the degeneracy (spin, isospin) of species $i$,
and $m_i$ its mass; the sum runs over all species. For simplicity, 
we assume for the moment $\mu=0$. Here
\be
\phi(m_i,T) = \int d^3p ~\exp\{\sqrt{p^2 + m_i^2}/T\}
\sim \exp-(m_i/T)
\ee
is the Boltzmann factor for species $i$, so that the ratio of the
production rates $N_i$ and $N_j$ for hadrons of species $i$ and $j$ is 
given by
\be
{N_i\over N_j}= {d_i \gamma_s^{\nu_i}\phi(m_i,T) 
\over d_j \gamma_s^{\nu_j}\phi(m_j,T)},
\ee
where $\nu_i=0,1,2,3$ specifies the number of strange quarks in species $i$.
We note that in the grand-canonical formulation the volume cancels out in
the form for the relative abundances.

\medskip

\begin{figure}[h]
\centerline{\psfig{file=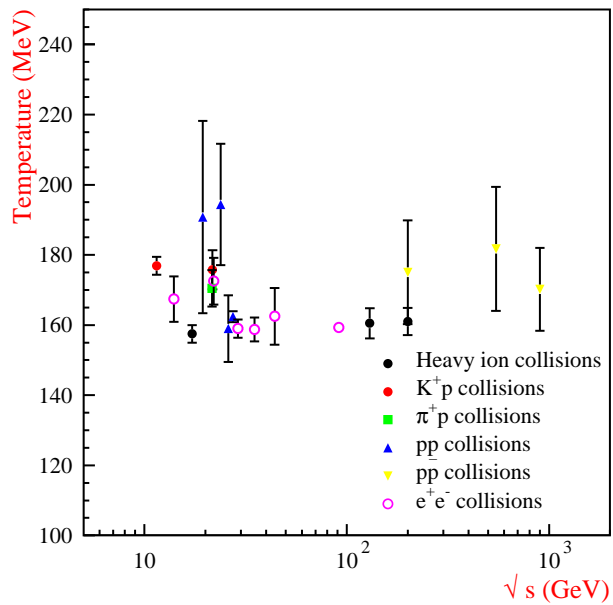,width=9cm}}
\caption{Hadronization temperatures for different collision energies and
different collision configurations \cite{becaT}.}
\label{ut1}
\end{figure}

Both temperature $T$ and strangeness suppression factor $\gamma_s$ have
been measured at various collision energies and for different collision
configurations. The resulting temperature of the emerging resonance gas is 
found to have  a universal value $T_c \simeq 160 \pm 10$ MeV for all (high) 
collision energies and all collision configurations, see Fig. \ref{ut1}.

\vskip-0.3cm

\begin{figure}[h]
\centerline{\psfig{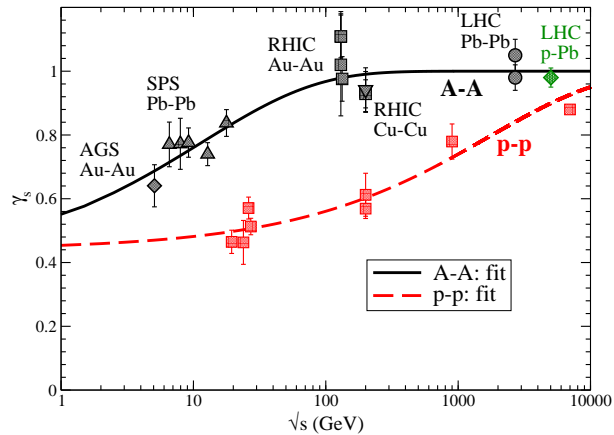}}
\caption{Strangeness suppression in $pp$ and $AA$
collisions at cms energy $\sqrt s$ \cite{data}.}
\label{ut2}
\end{figure}

In contrast, the strangeness suppression factor $\gamma_s$ depends over
a wide range on the collision energy and configuration; in particular, it
differs considerably in $pp$ and $AA$ interactions, see Fig. \ref{ut2}.
We note here that for $pp$ data, strangeness conservation is
generally taken into account exactly, rather than in a grand canonical
scheme with a strangeness chemical potential.

\vskip0.2cm 

The curves in Fig. \ref{ut2} are interpolation fits, leading to
\be
\gamma_s^p(s) = 1 - c_p \exp{(-d_p s^{1/4}}),
\label{ut3}
\ee
for $pp$ and
\be
\gamma_s^A(s) = 1 - c_A \exp{(-d_A \sqrt{A \sqrt s})}
\label{ut4}
\ee

\vskip0.3cm  
for $AA$ collisions, with$c_p=0.5595;~d_p=0.0242;~~c_A=0.606,~d_A=0.0209$.
It is evident from these fits that for sufficiently high collision energies,
one expects $\gamma_s$ to converge to unity for $pp$ as well as for $AA$
collisions, so that strangeness suppression then also disappears in
proton-proton interactions. This restoration of an equilibrium
grand-canonical strangeness
distribution has also been suggested on theoretical grounds \cite{CS}. 
 
\medskip

The results shown in Fig.\ \ref{ut2} suggest an obvious question: is there
a unified description for the strangeness suppression observed in $AA$ and
as well as in $pp$ collisions? Our aim here is to that if we replace  
the collision energy $\sqrt s$ by more natural thermal variables, then 
$\gamma_s$ in fact becomes one universal function. 

\medskip

\begin{figure}[h]
\centerline{\psfig{file=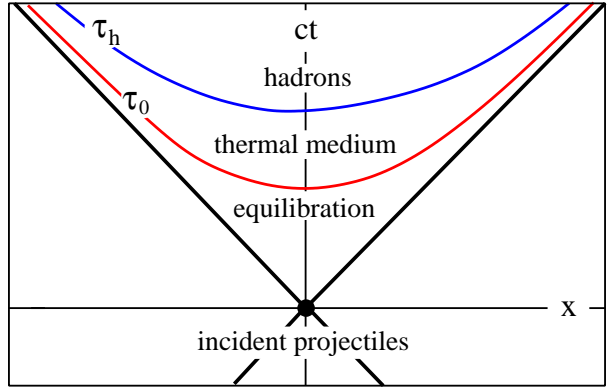,width=8cm}}
\caption{High energy collision in terms of longitudinal space $x$ and
time $t$ in the center of mass.}
\label{ut4a}
\end{figure}

The canonical view of a high energy collision is shown in Fig. \ref{ut4a}.
At proper time $\tau_0$, a thermal medium is formed, which at proper time
$\tau_h$ hadronizes, i.e., freezes out into an ideal resonance gas.

\medskip

\begin{figure}[h]
\centerline{\psfig{file=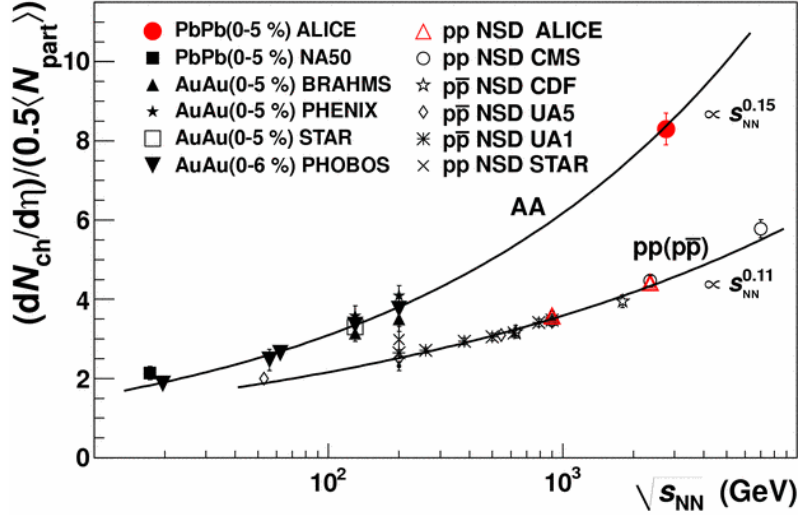,width=10.5cm}}
\caption{Average multiplicities in $pp$ and $AA$ collisions as function of 
cms energy $\sqrt s$ \cite{mult}.}
\label{ut6}
\end{figure}

The initial entropy density $s_0$ at the thermalisation time $\tau_0$ is for
one-dimensional hydrodynamic expansion given by the Bjorken relation

\be
s_0 ~\!\tau_0 \simeq
{1.5 A^x \over \pi R_x^2} \left({dN \over dy}
\right)_{y=0}^{x},~{\rm with}~ x \sim pp, pA, AA,
\label{ut5}
\ee

where $(dN/dy)^x(s)$ denotes the average multiplicity per unit central 
rapidity in the
corresponding reaction $x$ at collision energy $\sqrt s$, while $R_x$ gives 
the radius of the associated transverse area. This multiplicity has been
measured at different collision energies for $pp$ as well as for $AA$
collisions, with the result shown in Fig. \ref{ut6}

\medskip

The fit curves shown in this figure are given by

\be
\left({dN \over dy}\right)_{y=0}^{AA} = a_A(\sqrt s)^{0.3} 
+ b_A
\ee

and

\be
\left({dN \over dy}\right)_{y=0}^{pp} = a_p(\sqrt s)^{0.22} + 
b_p
\ee

\begin{figure}[h]
\centerline{\psfig{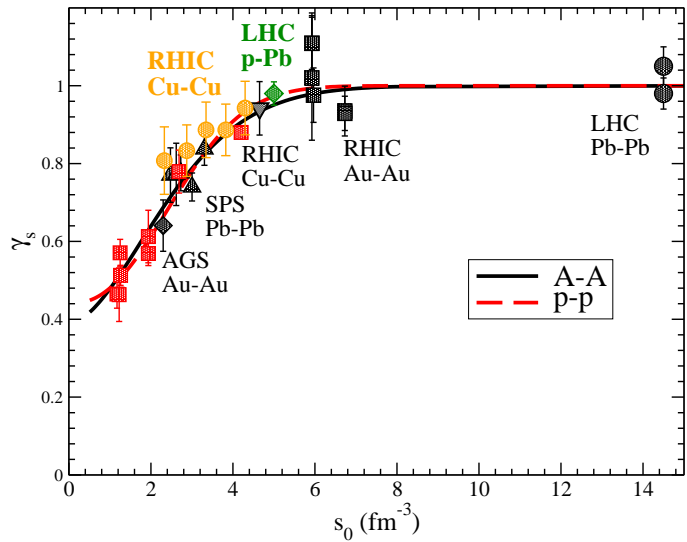}}
\caption{Strangeness suppression $\gamma_s(s_0)$ as function of the initial
entropy density $s_0$.}
\label{ut7}
\end{figure}

\vskip0.3cm
with $a_A=0.7613,~b_A=0.0534;~~a_p=0.797;~b_p=0.04123$. We can use 
these results to obtain the initial entropy density $s_0(s)$ in eq.\ 
(\ref{ut5}) as function of the collision energy. Next we then use 
eq's. (\ref{ut4}) and (\ref{ut5}) to express the strangeness suppression 
factor $\gamma_s$ as function of the initial entropy density of the 
medium produced in the corresponding collision. For the relevant parameters
in eq.\ (\ref{ut5}) we take $R_{pp}=0.8$ fm for $pp$, $R_{pPb}=R_{pp}
(0.5 {\bar N}_{part})^{1/3}$ with ${\bar N}_{part}=8$ for $pPb$
\cite{Abelev}, and $R_{AA}=1.25 A^{1/3}$ fm for $AA$.
The result thus obtained is shown in Fig. \ref{ut7}.


\vskip0.3cm

We see that we now indeed obtain a universal form of strangeness suppression,
in which the $pp$ and $AA$ interpolation curves coincide and with which
$pp$, $pA$ and $AA$ data agree. The relevant variable for the amount of 
strangeness suppression is thus the initial entropy density. As already 
indicated above, the pronounced strangeness suppression observed in $pp$ 
collisions at lower entropy densities disppears when collision energy and 
hence $s_0$ increase. Such an effect had been predicted some time ago as a 
consequence of the increasing life-time of the thermal medium with increasing 
collision energy \cite{CS,CPS}. 

\medskip

\begin{figure}[h]
\vskip0.5cm
\centerline{\psfig{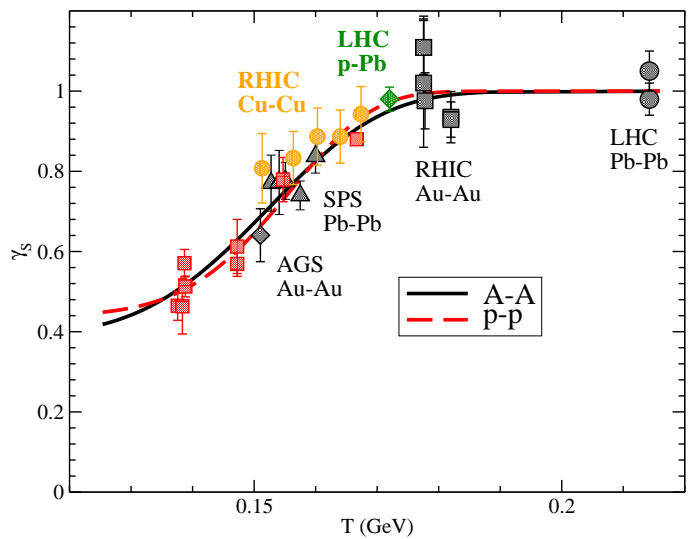}}
\caption{Strangeness suppression $\gamma_s(T)$ as function of the initial
temperature $T$.}
\label{ut8}
\end{figure}

Since finite temperature lattice QCD calculations \cite{Baza} provide us with 
the dependence of $s_0$ on the initial temperature $T$,  we can use the
the results of fig.\ \ref{ut7} to express $\gamma_s$ as a function of $T$. 
The resulting universal strangeness suppresson as function of the initial
temperature is shown in Fig. \ref{ut8}. Here we should keep in mind that
the lattice results hold for small values of the baryochemical potential
$\mu$, so that the $AA$ points at lowest temperatures (AGS) are to be
considered estimates only.

\medskip

From Fig. \ref{ut8} it is evident that if the initial
temperature of the thermal medium produced in the collision lies below the
deconfinement temperature, i.e., if the medium is in a hadronic state, there is
strangeness suppression, with $\gamma_s \simeq 0.5$. For media of initial
temperatures above the transition region, such a suppression is no longer
present; it disappears in a very narrow temperature range around $T_c \simeq
160$ MeV, for $pp$ as well as for nuclear collisions. In other words,
strangeness suppression disappears with deconfinement.The same relation to
the confinement/deconfinement point is of course also present in Fig. 
\ref{ut7}: here the transition occurs in a narrow band around the
corresponding entropy density at the transition point, 
$s_0 \simeq 3~{\rm fm}^{-3}$, again in accord with lattice results. 

\medskip

To obtain the universal form of strangeness suppression,
we had used the average multiplicities.
We could, however, use the above formalism also to derive
predictions for $\gamma_s(s)$ at a fixed cms energy as function of
the associated multiplicity. In particular, we could in this way
obtain the multiplicity dependence of strangeness suppression in $pp$ 
collisions at the LHC. This, however, requires as input the transverse

area, which also depends on the corresponding final multiplicity, since
high multiplicity events arise from large transverse area fluctuations. 
This has been estimated in specific approaches,
such as the color glass condensate \cite{trans-area}, but as such 
remains model-dependent. We therefore postpone such an analysis to
later work.

\medskip

In conclusion, we have shown that in terms of thermal variables,
strangeness suppression is a universal
phenomenon in all high energy hadronic interactions, $pp,~pA$ and $AA$.
It decreases with increasing initial entropy density of the produced thermal 
medium, or equivalently, with increasing initial temperature, and it 
vanishes with the onset of deconfinement. Above the deconfinement transition
regime, full equilibration of strange as well as non-strange secondaries
is reached for all high energy multihadron production.

\end{document}